# On the biological and cultural evolution of shame: Using internet search tools to weight values in many cultures

**Klaus Jaffe, Astrid Flórez, Cristina M Gomes, Daniel Rodríguez, Carla Achury**

Laboratorio de Evolución, Universidad Simón Bolívar, Caracas, Venezuela

**Abstract:** Shame has clear biological roots and its precise form of expression affects social cohesion and cultural characteristics. Here we explore the relative importance between shame and guilt by using Google Translate to produce translation for the words *shame, guilt, pain, embarrassment* and *fear* to the 64 languages covered. We also explore the meanings of these concepts among the Yanomami, a horticulturist hunter-gatherer tribe in the Orinoquia. Results show that societies previously described as "guilt societies" have more words for guilt than for shame, but the large majority, including the societies previously described as "shame societies", have more words for shame than for guilt. Results are consistent with evolutionary models of shame which predict a wide scatter in the relative importance between guilt and shame, suggesting that cultural evolution of shame has continued the work of biological evolution, and that neither provides a strong adaptive advantage to either shame or guilt. We propose that the study of shame will improve our understanding of the interaction between biological and cultural evolution in the evolution of cognition and emotions.

**Sobre la evolución biológica y cultural de la vergüenza: El uso de herramientas de búsqueda en Internet para una valoración relativa de estos conceptos en múltiples culturas**

**Resumen:** La vergüenza tiene claramente raíces biológicas y su forma concreta de expresión afecta a la cohesión social y otras características culturales. Aquí se explora la importancia relativa entre los conceptos "vergüenza" y "culpa" mediante el uso de Google Translate para buscar sinónimos de la palabra vergüenza, culpa, dolor y miedo en 64 idiomas. También exploramos los significados de estos conceptos entre los yanomami, una tribu de cazadores-recolectores en la Orinoquia. Los resultados muestran que las sociedades anteriormente descritas como "sociedades de culpa" tienen más palabras para la culpa que para la vergüenza, pero la gran mayoría, incluyendo las sociedades anteriormente descritas como "sociedades de vergüenza", tienen más palabras para la vergüenza que para la culpa. Los resultados son congruentes con modelos evolutivos de la vergüenza que predicen una gran varianza en la importancia relativa entre la culpa y la vergüenza en diferentes culturas; y sugieren que la evolución cultural de la vergüenza ha continuado el trabajo de la evolución biológica, basado en las ventajas adaptativas relativas de los sentimientos de vergüenza y culpa. Se identifica al estudio del sentimiento o emoción de la vergüenza como un modelo adecuado para comprender la interacción entre la evolución biológica y cultural de aspectos cognitivos y emocionales de nuestra conducta.

## Introduction

Psychologists base their research on proving theories. In natural sciences, when very little is known about the subject of study, exploratory research is preferred. As knowledge about the evolutionary origin of shame and guilt is very poor, we decided to explore some basic assumption about the existence of adaptive forces, be they biological or cultural in origin, which might have selected for a balance between the feelings of shame and guilt.

***The emotion of shame:***

Shame has been related to emotions and cognition and extensively studied in different contexts (11, 25, for example). The roots of the word *shame* are thought to derive from an older word meaning "to cover". The emotion of shame has clear physiological consequences. Its facial and corporal expression is a human universal, as was recognized already by Darwin (5). Looking away, reddening of the face, sinking the head, obstructing direct view, hiding the face and downing the eyelids, are the unequivocal expressions signaling shame. Shame might be an emotion specific to humans, as no clear description of it is known for animals. Behaviors in animals that share some aspects of shame include submission, both in interactions between adult-offspring and in other social contexts, and cryptic expressions of fear that are an attempt to hide when in presence of stronger rivals or potential predators. On the other hand, shame has been postulated as a cement of human societies upon which they can build and expand. Classical Greek philosophers, such as Aristotle, explicitly mention shame as a key element in building society.

Guilt is the emotion of being responsible for the commission of an offense, however, it seems to be distinct from shame. Guilt says “what I *did* was not good”, whereas shame says “I *am* no good" (2). For Benedict (1), shame is a violation of cultural or social values, while guilt feelings arise from violations of one's internal values. As proposed by Wikipedia, reflecting more popular notions of these concepts, “in a shame society the primary device for gaining control over children and maintaining social order is the inculcation of shame and the complementary threat of ostracism”. In contrast, in guilt societies control is maintained by creating and continually reinforcing the feeling of guilt (and the expectation of punishment now or in the hereafter) for certain condemned behaviors (1,7,9). Cultural differences between shame and guilt show both a high degree of universality of differential emotion patterning, and important cultural differences in emotion elicitation, regulation, symbolic representation, and social sharing (38). Shame and guilt have often been used as a classifier for cultures.

The emotion of shame has suffered both biological and cultural evolution. Although the mechanisms upon which both types of evolution are based on differ widely, its dynamics seem to be very similar (23, 30). A deeper understanding of shame can serve as an empirical test of the relation between cultural and biological evolution. Cultural evolution could continue the work of biological evolution, if the adaptive advantages of shame are maintained over time; or both types of evolution might drive the behavioral-cultural system to different outcomes, if the adaptive advantages of shame differ in different cultures.

***The biological view of shame:***

Ethologically, shame is related to unconscious physiological reactions and therefore can be viewed as an honest signal regulating social interactions, where the benefits to society is the identification of trustworthy individuals and the benefit to the individual is reducing or avoiding social punishment when transgressing a social norm. The uncontrollable trigger of shame signals makes it an honest signal trait which serves to regulate compliance with social norms. In economy, reliable “public information” has to be readily apparent, which seems to be the case for shame. Unobservable emotions such as guilt may be of value to the receiver but constitutes in economy “private information”. Thus, in economic and biological terms, adaptive pressures acting upon the evolution of shame differ from those acting on that of guilt.

Shame has evolutionary advantages to both individual and society, but the lack of shame also has evolutionary advantages as it allows cheating and thus benefiting from public goods without paying the costs of its build up. These divergent adaptive forces have an interesting effect on the evolutionary dynamics of shame. Computer simulations of virtual societies formed by agents who can cooperate synergistically, mutualistically or egoistically showed that genes or memes coding for shame will always evolve and eventually displace genes or memes coding for shamelessness, but the opposite will also happen. The outcome of computer simulations is a dynamic unstable equilibrium between both strategies, producing as a result chaotic-like dynamics with spells of stable shameful populations followed by spells of shameless populations (21). These results thus predict variable levels of shame among distinct populations.

If the adaptive forces leading humans to evolve a common expression of shame are the same as those favoring shame in contemporary human societies, we should expect that all human populations will possess genes allowing for emotions and elicitors of shame but express them in different ways. If guilt is also a universal emotion, the equilibrium between the emotions of guilt and shame, should vary, irrespective of the possibility that guilt alone is an evolutionary stable or unstable emotion. If the simulations just mentioned, however, do not reflect reality, other outcomes are possible.

Adaptation seems to have selected for intermediate levels of shame among humans. Excess of shame has been associated to pathologies. For example, high levels of shame are linked to mental illness in the USA (10,28, 37, 42) and to physiological stress (6, 13, 32). On the other hand, a deficit of shame also relates to pathological states, as shameless individuals are prone to suffer psychopathic syndromes (3, 18, 33).

***Shame in different cultures:***

Dodds (7) coined the distinction between guilt and shame cultures and postulated that in Greek cultural history, shame as a social value was displaced, at least in part, by guilt in guiding moral behavior. Hiebert (15) differentiates between guilt and shame societies as follows: "Guilt is a feeling that arises when we violate the absolute standards of morality within us, when we violate our conscience. A person may suffer from guilt although no one else knows of his or her misdeed; this feeling of guilt is relieved by confessing the misdeed and making restitution. True guilt cultures rely on an internalized conviction of sin as the enforcer of good behavior, not, as shame cultures do, on external sanctions. Guilt cultures emphasize punishment and forgiveness as ways of restoring the moral order; shame cultures stress self-denial and humility as ways of restoring the social order". Thus, two distinct cultures seem to exist based on distinct emphasis or importance of the relation between guilt and shame.

In their extensive review of cultural models of shame and guilt, Wong and Tsai (45) make an important effort to incorporate the available empirical findings to the discussion of the cultural relevance of shame and guilt. They propose an individualistic and a collectivistic model of shame and guilt, where the valuation, elicitors and behavioral consequences, as well as the distinction between shame and guilt, vary systematically across individualistic and collectivistic cultures. Their review includes a relatively small group of cultures and they convincingly argue for the need of further cross cultural research.

Cross cultural studies of emotions have been reviewed before (4, 16, 26, 27, 36, 40, 41, 43, 45), but do not focus on shame and will not be discussed further here.

**Using crowd-sourcing and automated search techniques in scientific research**

Comparing anthropological and linguistic studies of different cultures introduces an important amount of variations not related to the cultural differences, but depend on the focus, motivation and professional experience of the diverse researchers, Thus, these studies are not appropriate for our aim here. The anthropological and social studies we found are limited to a few social groups or cultures. A broad study covering a large amount of different cultures is required for our aims here. Therefore, we used an approach, based on automatic algorithms which minimize subjective criteria, that allows covering a large number of cultures and thus enables us to make broad inter-cultural comparisons. Thanks to the fast advance of information technology, such approaches are now possible. That is, we used the same method with the same errors to analyze with minimal effort a large set of different languages to explore conceptual relationships between shame and guilt and some related concepts. By studying the number of words used for shame, guilt, embarrassment, pain and fear, and in some cases the contexts in which these are used, we aim at gaining insight into the universality of the relationship between shame and guilt, and obtain a crude estimate of its relative importance as assed through the subtle variations in describing these feelings in each culture.

Many scholars avoid using Wikipedia or Google search tools as they consider them not to be scientifically validated and recognized methods. We believe that these modern tools have many benefits over traditional one. They are cheap to use, easily replicable, cover a very broad area of search space and are at least as reliable as traditional research methods. For example, Wikipedia is less error prone than Encyclopedia Britannica (12, 17); and Google Translate is as accurate as more traditional methods (35). Here we use Google translate as a research tool, as it provides the broadest access to concepts in the largest number of languages available using the same algorithm. This does not imply that our method might commit other types of errors, but the contrast of different research efforts allows to advance science.

# Methods

We wanted to assess the relative importance of the concept shame and guilt in different cultures. This can be done by counting the number of synonyms of each concept in each language and assuming that cultures that have developed more synonyms for a concept have dedicated more attention to it and thus are likely to give it more importance than other concept which have fewer or no synonyms.  For example Inuit's use dozen of words for snow whereas Yanomami have none. This difference clearly reflects the relative importance of snow in each culture. Inuit's contact snow nearly all year long whereas Yanomami never see it. Using "Google Translate" we translated and noted the synonyms of the English words: *shame, guilt, pain, embarrassment* and *fear* to the 64 languages covered (Table 1). For the data on English, Yanomami and Sanema, we translated from the Spanish words: *vergüenza, culpa, dolor, pena* and *miedo.* Google Translate, has an uneven coverage of the languages reported. For each language, however, it is likely to have similar coverage of the concepts studied. Thus, any difference in coverage of a language less likely to affect the ratio of synonyms of two related concepts. But important errors have been detected. For example Czech and Slovak languages showed large differences in the number of terms for guilt and shame (1 & 3 for Czech and 5 & 9 for Slovak) given that the two languages are very closely related (Štěpán Bahník personal communication). The ratio between the numbers, however, showed less variation (0.3 and 0.5). This error is somehow captured by the 95% standard error interval in Figure 2, where Slovak is positioned below the upper 95% limit. Google translate produces more erroneous data. For example it reports a word in Slovak to describe indistinctly shame and pain; this is not true as the word *otrava* means either poisoning, nuisance or boredom. Correcting this error in the figure, changing the color of Slk, improves the picture but more errors with other languages are to be expected. Thus, the analysis performed here will have to be repeated in the future when better data will become available.

Data for the Sanema, an indigenous horticulturist group belonging to the cultural-ethnic complex of the Yanomami (31), was collected in communities located on the banks of the Caura River, Bolivar State, Venezuela, during 2012. With the help of native Spanish-Sanema translators, we interviewed 16 Sanema people, members of the Ikutu community. We asked them questions regarding words used in different contexts and contexts in which different words are used. Thus, we were able to determine which words are commonly used in contexts usually related to shame, guilt, pain, embarrassment and fear. For Yanomami we used the Spanish-Yanomami dictionary by Marie Claude Mattei-Muller (31).

## Results

Results presented in Figure 1 show that all 64 languages studied have at least one concept related to *Shame* and one to *Guilt* and most languages have more words for *Shame* than for *Guilt*. In Figure 1, languages which have at least one word which describes indistinctly *shame* and *embarrassment* are indicated in red. The total list includes Arabic, Azerbaijani, Catalan, Croatian, Filipino, French, Irish, Macedonian,

Malay, Persian, Swahili, Turkish, Slovak, and Spanish. Slovak has a word to describe indistinctly *shame* and *pain*; and Yanomami and Sanema use the same word for *shame, embarrassment* and *fear.*

The line in Figure 2 represents the regression with its 95% confidence interval. This regression is close to a proportional line of two words for shame for each word for guilt. Dots below the line indicate languages where more words than the 2/1 proportion are used for shame than for guilt. Interestingly, dots in red (i.e. languages which confuse shame with another concept) are found only below the upper 95% confidence limit of this regression line.

As deduced from Figure 1, 49 languages have clear linguistic separation between *Shame* and the four concepts explored (*Guilt, Pain, Embarrassment, Fear*); 14 languages confuse *Embarrassmen*t with *Shame.* Yanomami has a diffuse relationship between *shame, embarrassment* and *fear.* The linguistic separation is especially conspicuous in languages with many words for *shame* and *guilt*, such as Hebrew, Latin and Bengali.

We might underestimate the overlap of the meaning of words. For example, in the case of Chinese, no overlap between the five concepts is reported using Google Translate in Figure 1. Yet, linguistic-conceptual studies of guilt and shame revealed an important overlap between several of these concepts in Chinese (29). The authors found, at the highest abstract level, two large distinctions of “shame state, self-focus” and “reactions to shame, other-focus.” While the former describes various aspects of actual shame experience that focuses on the self, the latter focuses on consequences of and reactions to shame directed at others. Shame state with self-focus contained three further sets of meanings: 1- one's fear of losing face, 2- the feeling state after one's face has been lost, and 3- guilt. Reactions to shame with other-focus also consisted of three further sets of subcomponents at the same level: 4- disgrace, 5- shamelessness and its condemnation, and 6- embarrassment. Except for guilt, there were several sub-clusters under each of these categories.

Our results using Google Translate show no overlap between *Guilt* and *Shame* in any of the languages studied. This is interesting in the light of the distinction made previously between *Shame* and *Guilt* societies by Wong & Tsai (45). Although some cultures are multi-linguistic it is not difficult to match the languages analyzed with the societies reported by Wong & Tsai (45). There seems to be a significant overlap between both ways of measuring the shame-guilt dimension. All societies that had been reported as "guilt societies" speak languages that have more words for guilt as that predicted by the regression in Figure 2 (i.e. less than 2 synonyms for shame for each synonym of guilt), i.e. Anglo-Saxon western societies; whereas the "shame societies", reported in the literature (16, 19, 39): Japan, Persia, Arabs and China, use languages with two or more words for shame than for each synonym for guilt.

Yanomami is very different from other languages in not having clear divisions between several of the concepts studied. Yanomami´s are a very distinct amerindian hunter-gather community in the Orinoco-Amazon basin of Brazil and Venezuela, and differ from other amerindians and western cultures in that they spend more time in communal relationships so that they are focused more on society than on the individual (8). Their knowledge of nature is more restricted than that of other communities living in the same area (14, 24). Yanomami is the only language that uses the same word to refer to shame, pain, fear, guilt and embarrassment. We thus, studied in more depth the concept of shame among Sanema or Sanöma, one of the four dialects of Yanomami. The word most Sanema related to shame was “kili”. Examples of the context when they feel “kili” are: a tiger appears in the forest; you kill somebody from another community; your daughter is going to die; everybody looks at your underwear; you are caught stealing; you soil your pants while among others; a doctor gives you an injection; you hit your wife and others find out; you are unfaithful to your husband and others find out; you are going to be hit with a machete.

Linguistic families do not aggregate according to the relationship of the number of synonyms for shame and guilt (Figure 3). For example, Latin languages (Latin, Spanish, Romanian, Italian, French, Portuguese) differ widely in their relative importance between guilt and shame and appear scattered over all area in Figure 3. Similarly, Hebrew and Yiddish; or Korean, Japanese and Chinese; which share important aspects of written language; etc. are separated by the principal component analysis in Figure 3. The linguistic families examined did not cluster according to the relative importance they give to shame and guilt.

Testing for a general validity of Dodds proposition that older shame-cultures may evolve towards guilt-cultures, as shown in Greek literature, we compared the ratios of the number of synonyms for shame and guilt in Latin with Italian. The ratios are 0.89 and 2.5 respectively, meaning a historical transition from guilt-culture in Latin to shame-culture in Italian, suggesting a historical development that is inverse to that suggested by Dodds for ancient to classical Greek.

Languages with extreme Shame/Guilt ratios in relative number of synonyms (S/G) are: S/G < 1 English, Hungarian, Telugu, Bengali, Latin; S/G > 3 Hebrew, Italian, Korean, Yiddish, Spanish, Lithuanian.

## Discussion

Our study supports the use of Google Translates to compare the relative "importance" of different concepts in different cultures. Here we showed that all the 64 languages examined, have a unambiguous word for shame and guilt and sharply distinguish between them. This finding is in agreement with the view that there is a high degree of universality in the different emotional patterning and in the cultural differences in emotional elicitation of shame and guilt (38). This universality, however, does not preclude divergence in the importance of shame in different societies. The diversity of linguistic usage for shame and guilt also suggests that the cultural evolution of shame has continued the work of biological evolution. Our results showed a wide scatter in the relative importance or dept of naming subtle differences between guilt and shame, as estimated by quantifying the number of synonyms produced by Google Translate of each concept in each language. Despite this scatter, and independent of language families, all societies or cultures that had been referred to as "shame societies" in the literature had high scores on relative frequency of words for shame/guilt, whereas those referred to as "guilt societies" had a low score in this relationship. The present study provides for testable predictions as it suggests which society should be closer to a "guilt" or "shame society”, based on their language, which can be confirmed or negated with further anthropological or cultural studies. For example, Hungarian, Telugu, Bengali, Latin should be spoken in “guilt” societies; whereas Hebrew, Italian, Korean, Yiddish, Spanish and Lithuanian should be spoken in “shame societies”.

Results are consistent with evolutionary models of shame which predict a wide scatter in the relative importance between guilt and shame. Neither biological nor cultural evolution provides a strong adaptive advantage to either shame or guilt. The divergence between guilt and shame societies seems to be a natural outcome of the distinct adaptive advantages of shame and guilt, as predicted from simulating shame in virtual societies (21). These simulations showed that shame, together with pro-social punishment and social cooperation, produce fluctuating dynamics of social cooperation, achieving long periods where the populations stabilizes pro-social behavior interspersed with periods where selfish behavior predominates. Although shamelessness could in theory out-evolve shamefulness, empirical evidence suggest otherwise. There is overwhelming evidence that cooperation is often more successful in evolution than confrontation (see 20, 22, 34, 44, for example) suggesting that shamelessness, good for confrontation, is not likely to out-evolve shamefulness which is favors cooperation.

The data presented here seems to be consistent with this view. Some societies place more importance on guilt than on shame, but the large majority does the inverse. A few societies have a concept of shame that is indistinguishable from fear, embarrassment or guilt, whereas others separate these concepts very clearly. But all societies know what shame is when they see it.

We considered this work to be a preliminary exploration that contributes to open new windows into the search for the evolution of emotions. The study of shame and guilt offers a good access to study the interaction between biological and cultural evolution.  Few cognitive features are so related to our social instinct as shame, thus, it is astonishing that we know so little about shame. More extensive interdisciplinary analyses including linguistic studies, finer anthropological synthesis of the literature, neuroethology and other disciplines, should help improve our insight into the cognition behind emotions and its evolution.

**Acknowledgements**. We thank anonymous referees for helpful detailed comments and the National Science Foundation and the Leakey Foundation for financial support for the work with the Sanema and the Fundación La Salle for access to their library.

**Figure 1**. Synonymous as reported by Google Translate in 2013

| | **shame** | **guilt** | **pain** | **embarrassment** | **fear** |
|---|---|---|---|---|---|
| **Afrikaans** | skande | skuld | pyn | verleentheid | vrees |
| **Albanian** | turp<br>turpërim<br>çnderim | faj<br>veprim<br>fajshëm | dhimbje<br>dhembje<br>mundim<br>vuajtje<br>dëshpërim<br>ofshamë<br>ndëshkim | siklet<br>vështirësi<br>telash<br>telendisje<br>zor<br>çoroditje<br>turbullim<br>ngatërrim | frikë<br>rrezik<br>shqetësim |
| **Arabic** | عار<br>الخجل<br>خجل<br>خزي<br>حياء<br>هوان<br>مصدر خزي<br>عار خزي<br>ارتباك | م<br>تهمة<br>معصية<br>ذنب إثم | ألم<br>الم<br>وجع<br>عناء<br>حزن<br>أسى<br>عقوبة<br>جهد | ارتباك<br>حيرة<br>عائق<br>مشاكل مالية | خوف<br>خشية<br>خطر |
| **Armenian** | ամոթ<br>խայտառակու<br>թյուն<br>ամոթի<br>զգացում<br>անպատվությ<br>ուն | Մեղք<br>հանցանք | ցավ<br>վիշտ<br>ճիգ<br>ջանք<br>պատիժ<br>տանջանք | դժվարություն<br>ծանր<br>դրություն<br>շվարում<br>շփոթություն | վախ<br>ահ<br>անհանգստութ<br>յուն<br>երկյուղ<br>կասկած |
| **Azerbaijani** | ayıb<br>rüsvayçılıq<br>rüsvay<br>xəcalət<br>ar<br>eyib<br>xar<br>abır<br>bədnamlıq<br>həya<br>biabırçılıq<br>eyiblik<br>ayıblıq | təqsir<br>günah<br>günahkarlıq<br>səbəbkarlıq<br>suç<br>babal | ağrı<br>gizilti<br>ağrı-acı | çaşqınlıq<br>sıxılma<br>xəcalət<br>qısıntı<br>tutulma<br>çətinlik<br>xəcalətlilik | qorxu<br>vahimə<br>zaval<br>heybət<br>həzər<br>qorxu-hürkü<br>vahimətilik<br>zəhm<br>xof |
| **Basque** | lotsa | erruduntasuna | mina | horretaz | beldurra |
| **Belarusian** | ганьба | віна | боль | збянтэжанасць | страх |
| **Bengali** | লজ্জা<br>অপমান<br>শরম<br>ত্রপা<br>গ্লানি<br>কলঙ্ক | অপরাধ<br>দোষ<br>অপরাধিত্ব<br>পাপিষ্ঠতা<br>পাপ<br>কলুষ<br>কল্মষ<br>কু | ব্যথা<br>ব্যাথা<br>বেদনা<br>কষ্ট<br>শাস্তি<br>তকলিফ<br>শূল<br>শূলানি<br>মানসিক যন্ত্রণা<br>শারীরিক যন্ত্রণা<br>পীড়া<br>মর্মযন্ত্রণা<br>দৈহিক বেদনা<br>আয়াস<br>পেন<br>ক্লেশ<br>ক্ষোভ<br>আর্তি<br>কামড় | বিব্রত অবস্থা<br>থতমত<br>মানসিক বিহ্বলতা<br>বিব্রত করণ<br>দেনাগ্রস্ত অবস্থা<br>বিবৃতকারী বস্তু<br>বিভ্রান্তি<br>বিব্রত হইবার কারণ<br>বিমূঢ়তা<br>নাকাল<br>অভিভব<br>বিহ্বলতা<br>বৈক্লব্য<br>অভিভাব<br>হতবুদ্ধিতা | ভয়<br>ভীতি<br>আশঙ্কা<br>আতঙ্ক<br>শঙ্কা<br>পরোয়া<br>ডর<br>দর<br>কুণ্ঠা<br>অভিশঙ্কা<br>ত্রাস<br>অধীরতা<br>খতরা |
| **Bulgarian** | срам<br>позор<br>безобразие<br>свян<br>неприятност | вина<br>грях<br>закононарушение | болка<br>мъка<br>страдание<br>огорчение<br>наказание | смущение<br>объркване<br>затруднение<br>пречка<br>неудобно<br>положение | страх<br>опасение<br>ужас<br>опасност<br>вероятност |
| **Catalan** | vergonya | culpa | dolor | vergonya | por |
| **Chinese Simple** | 耻辱<br>羞耻<br>羞辱<br>羞<br>耻<br>侮辱 | 有罪<br>罪<br>辜 | 疼痛<br>痛苦<br>痛<br>疼<br>酸痛<br>苦<br>苦痛<br>酸疼<br>酸<br>楚 | 困窘<br>困难<br>阻碍<br>爽然 | 怕<br>忧<br>怵<br>忧心<br>忌<br>惮<br>懔<br>惶<br>怖 |
| **Chinese Traditional** | 恥辱<br>羞恥<br>羞辱<br>羞<br>恥<br>侮辱 | 有罪<br>罪<br>辜 | 疼痛<br>痛苦<br>痛<br>疼<br>酸痛<br>苦<br>苦痛<br>酸疼<br>酸<br>楚 | 困窘<br>困難<br>阻礙<br>爽然 | 怕<br>憂<br>怵<br>憂心<br>忌<br>憚<br>懍<br>惶<br>怖 |
| **Croatian** | sramota<br>šteta<br>sram<br>stid<br>bruka | krivica<br>grijeh | Bol<br>patnja | stid<br>zbunjenost<br>zabuna | Strah<br>zort |
| **Czech** | hanba<br>ostuda<br>stud | vina | bolest<br>otrava<br>žal<br>trest<br>zármutek<br>hoře | rozpaky<br>rozpačitost | strach<br>obava<br>bázeň |
| **Danish** | skam | skyld | smerte | forlegenhed | frygt |
| **Dutch** | Schaamte<br>schande | Schuld<br>misdaad | pijn<br>smart<br>leed<br>lijden<br>zeer<br>straf<br>wee | verlegenheid<br>verwarring<br>hinder<br>moeilijkheid<br>benardheid<br>penarie<br>knelpunt | angst<br>vrees<br>ontzag<br>beklemming |
| **English** | **shame** | **guilt**<br>fault<br>blame<br>onus | **pain**<br>penalty<br>sorrow<br>trouble<br>distress<br>embarrassment<br>heartache<br>misery<br>infliction<br>labor<br>forfeit<br>fash | **embarrassment**<br>sultriness | **fear**<br>apprehension<br>trepidation<br>nervousness<br>fearnought |
| **Estonian** | häbi<br>kahju<br>häbistus<br>blamaaž | süü<br>süütunne | valu<br>piin<br>vaev | raskused<br>rahapuudus | hirm<br>kartus<br>pelg |
| **Filipino** | kahihiyan<br>pagkahiya<br>hiya<br>pagkapahiya<br>baho | pagkakasala<br>kasalanan | sakit<br>kirot<br>kalungkutan<br>lungkot<br>sama ng loob | kahihiyan<br>paghiya<br>pagkapahiya | takot<br>pagkatakot<br>pangamba<br>sindak<br>pagkasindak<br>alarma<br>bakla |
| **Finnish** | häpeä<br>Häpeä<br>häpeäntunne | syyllisyys | kipu<br>kivut<br>tuska<br>kärsimys<br>murhe<br>piina<br>riesa<br>kiusankappale | hämmennys<br>este<br>vaikeus<br>taloudelliset<br>vaikeudet<br>rahapula | pelko<br>huoli<br>levottomuus |
| **French** | honte<br>dommage<br>confusión | culpabilité<br>accusation | douleur<br>peine<br>souffrance<br>mal<br>tristesse<br>effort<br>punition | embarras<br>confusion | peur<br>crainte<br>danger |
| **Galician** | vergoña | culpa | dor | constrangimento | medo |
| **Georgian** | სირცხვილი | დანაშაული | ტკივილი | არეულობის | შიში |
| **German** | Scham<br>Schande<br>Schmach<br>Beschämung<br>Unwürdigkeit | Schuld<br>Täterschaft | Schmerz<br>Leid<br>Qual<br>Qualen<br>Kummer<br>Mühe | Verlegenheit<br>Peinlichkeit<br>Beschämung<br>Betroffenheit<br>Betretenheit | Angst<br>Furcht<br>Befürchtung<br>Scheu<br>Risiko<br>Respekt |
| **Greek** | Ντροπή<br>Όνειδος<br>εντροπή | ενοχή | πόνος<br>λύπη | αμηχανία | Φόβος ,φοβία |
| **Gujarati** | શરમાવવું<br>લાંછન<br>શરમની બાબત<br>કલંક | અપરાધ<br>દોષ<br>ગુનો | પીડા<br>દર્દ<br>વેદના<br>દુઃખ<br>કષ્ટ | અકળામણ<br>ગભરાટ<br>ઉચાટ<br>મૂંઝવણ | ભીતિ<br>ત્રાસ<br>ચિંતા ઈત્યાદિ<br>અનિષ્ટ ના ખ્યાલથી મનમાં<br>ઉદભવતી અમુક પ્રકારની<br>લાગણી ભય |
| **Haitian** | wont | koupab | doulè | anbarasman | pè |
| **Hebrew** | בושה<br>חרפה<br>כלימה<br>קלון<br>בזיון<br>בשת<br>שמצה<br>גנאי<br>בשת פנים<br>בוז<br>גנוי<br>קיקלון<br>דראון | אשמה<br>אשם<br>חטא<br>חובה | כאב<br>מכאוב<br>עצב<br>חבל<br>מחוש<br>כאיבה<br>דאבון<br>לבט<br>מזור | מבוכה | פחד<br>חשש<br>אימה<br>מורא<br>יראה<br>דאגה<br>אימתה<br>בעתה<br>דחילה<br>מגורה |
| **Hindi** | शर्म<br>लज्जा<br>लाज<br>अपमान<br>बदनामी<br>शर्मदगी<br>बेइज़्ज़ती<br>चिढ़ | अपराध<br>पाप<br>जुर्म<br>पातक<br>दुष्टता | दर्द<br>पीड़ा<br>कष्ट<br>वेदना<br>व्यथा<br>दंड<br>क्लेश<br>यंत्रणा | उलजाव<br>उलझेड़ा | डर<br>भय<br>आशंका<br>चिंता<br>झिझक<br>शंका<br>त्रास<br>संत्रास<br>ख़ौफ़<br>रोब<br>झझक<br>संवेग<br>खटका |
| **Hungarian** | szégyen<br>gyalázat | bűnösség<br>vétkesség<br>vétek | fájdalom<br>szenvedés<br>vajúdás | zavar<br>feszengés | félelem<br>aggódás |
| **Icelandic** | skömm | sekt | verkur | vandræði | ótti |
| **Indonesian** | malu<br>rasa malu<br>penghinaan<br>noda<br>keaiban<br>kecelaan<br>arang di muka<br>fadihat<br>sakit hati | kesalahan<br>perasaan bersalah<br>dosa | rasa sakit<br>sakit<br>kesakitan<br>rasa nyeri<br>penderitaan<br>kepedihan<br>perasaan sakit<br>hukuman | kejengahan<br>keadaan<br>memalukan | ketakutan<br>rasa takut<br>kecemasan<br>kekuatiran<br>kegentaran<br>kebimbangan<br>perasaan hormat dan takut<br>gentaran |
| **Irish** | náire | ciontacht | pian | náire | eagla |
| **Italian** | Vergogna<br>Peccato<br>Pudore<br>Onta<br>Disonore<br>Obbrobrio<br>indecenza | colpa<br>colpevolezza | dolore | imbarazzo | paura<br>timore<br>spavento |
| **Japanese** | 恥<br>恥辱<br>羞恥心<br>不名誉<br>辱<br>面汚し<br>不面目<br>ひどいこと<br>破廉恥<br>はじらい<br>赤恥<br>うらいこと | 有罪<br>罪状<br>罪科 | 痛み<br>疼痛<br>苦痛<br>苦しみ<br>傷み<br>悲哀<br>苦心<br>憂苦<br>憂悶<br>受難<br>憂患 | 当惑 | 恐れ<br>心配<br>懸念<br>虞<br>危惧<br>憂慮<br>畏怖<br>フェア<br>憂懼<br>疑心<br>不安心<br>怖じ気<br>畏懼<br>危虞<br>心労<br>気遣い<br>鬼胎<br>危懼<br>危疑<br>焦慮 |
| **Kannada** | ಅವಮಾನ<br>ನಾಚಿಕೆ<br>ಲಜ್ಜೆ | ಅಪರಾಧ<br>ತಪ್ಪಿತ<br>ಅಪರಾಧಿತ್ವ | ನೋವು | ಪೇಚು | ಹೆದರಿಕೆ<br>ದಿಗಿಲು<br>ಅಂಜಿಕೆ |
| **Korean** | 치욕<br>창피<br>부끄러움<br>잡된 행실<br>심한 짓<br>쓰라린 짓<br>분한 짓<br>우세 | 죄의식<br>죄 | 고통<br>통증<br>아픔<br>고생<br>국부적인 아픔<br>벌<br>통고 | 당황<br>난처<br>당황케하는 것<br>무안케하는 것<br>당황케하는 사<br>람<br>무안케하는 사<br>람<br>재정 | 무서움<br>근심<br>신에 대한 두려움 |
| **Latin** | verecundia<br>rubor<br>pudor<br>dedecus<br>probrum<br>turpe<br>flagitium<br>obprobrium | reatus<br>culpa<br>noxa<br>noxia<br>crimen<br>promeritum<br>reatitudo<br>peccatum<br>colpa | dolore<br>pena<br>sofferenza<br>male<br>fatica<br>travaglio<br>castigo<br>doglie del parto | confusioni | terror<br>formido<br>timor<br>pavor<br>trepidatio<br>metus<br>tremor<br>reverentia |
| **Latvian** | kauns<br>nepatika | vaina | sāpes | apmulsums | bailes |
| **Lithuanian** | Gėda<br>Nešlovė<br>Apmaudas<br>Akibrokštas<br>sarmata | kaltė | skausmas<br>kančia<br>gėla<br>sielvartas<br>širdgėla<br>bausmė<br>skaudulys | varžymasis<br>sutrikimas<br>keblumas<br>drovėjimasis<br>sumišimas<br>sunkumas | baimė<br>nuogąstavimas<br>būgštavimas<br>bijojimas<br>būkštavimas<br>galimybė<br>bailė |
| **Macedonian** | срам | вина | болка | срам | страв |
| **Malay** | malu | bersalah | kesakitan | malu | takut |
| **Maltese** | mistħija | ħtija | uġigħ | imbarazzament | biża |
| **Norwegian** | skam | skyld | smerte | forlegenhet | frykt |
| **Persian** | م<br>خجالت<br>ننگ<br>شرمساری<br>سرافکندگی<br>فضاحت<br>خجلت<br>ازرم<br>عار | گناه<br>جرم<br>تقصیر<br>مجرمیت<br>بزه | د<br>رنج<br>زحمت<br>تیر | خجالت<br>خجلت | س<br>هراس<br>وحشت<br>بیم<br>خوف<br>پروا |
| **Polish** | wstyd<br>hańba<br>zawstydzenie<br>kompromitacja<br>sromota<br>niesława | wina<br>przestępstwo<br>karygodność | ból<br>cierpienie<br>bolesność<br>boleść<br>męka<br>przykrość<br>dolegliwość<br>obolałość<br>strapienie | zakłopotanie<br>skrępowanie<br>zaaferowanie<br>ambaras | strach<br>lęk<br>bojaźń<br>trwoga<br>przestrach<br>obawianie się |
| **Portuguese** | vergonha<br>pudor<br>ignomínia<br>opróbrio<br>desonra | culpa<br>crime<br>delito | dor<br>sofrimento<br>pena<br>aflição<br>mágoa<br>esforço<br>pesar<br>trabalho<br>punição<br>castigo<br>dores de parto<br>abacaxí | embaraço<br>constrangimento<br>estorvo<br>dificuldade<br>perturbação<br>empecilho | medo<br>temor<br>receio<br>terror |
| **Romanian** | ruşine<br>ocară<br>pudoare<br>scandal<br>necinste<br>decădere | vinovăţie<br>vină<br>culpă<br>păcat<br>acuzare | durere<br>suferinţă<br>chin<br>eforturi<br>pedeapsă<br>efort<br>pacoste<br>necaz<br>oboseală<br>durerile facerii<br>supărare<br>ameninţare cu pedeapsă<br>bătaie de cap<br>calamitate | jenă<br>stinghereală<br>bucluc<br>belea<br>jenă financiară<br>strâmtorare<br>greutate | frică<br>teamă<br>temere<br>spaimă<br>îngrijorare<br>respect profund |
| **Russian** | стыд<br>позор<br>срам<br>бесславие<br>досада<br>неприятность | вина<br>виновность<br>чувство вины<br>грех<br>вин `a | боль<br>страдание<br>горе<br>огорчение | смущение<br>замешательство<br>затруднение<br>конфуз<br>запутанность<br>помеха<br>препятствие | страх<br>опасение<br>боязнь<br>вероятность<br>возможность |
| **Serbian** | срамота<br>штета<br>стид<br>срам<br>брука | кривица<br>грех | бол<br>патња | стид<br>збуњеност<br>забуна | страх<br>зорт |
| **Slovak** | hanba<br>škoda<br>potupa<br>stud<br>smola<br>nepríjemnosť<br>otrava<br>zneuctenie<br>škvrna | vina<br>pocit viny<br>trestuhodnosť<br>zodpovednosť<br>chyba | bolesť<br>utrpenie<br>trest<br>otrava<br>omyl<br>muka | rozpaky<br>ťažkosti<br>finančné ťažkosti | strach<br>obava<br>bázeň<br>nebezpečie |
| **Slovenian** | sramota | krivda | bolečina | zadrega | Strach |
| **Spanish** | vergüenza<br>lástima<br>oprobio<br>deshonra<br>mula | culpa | dolor<br>pena<br>sufrimiento | vergüenza<br>embarazo<br>bochorno<br>turbación<br>desconcierto<br>pena<br>estorbo<br>azoramiento<br>empacho<br>panza | miedo<br>temor<br>horror<br>aprensión |
| **Swahili** | aibu | hatia | maumivu | aibu | rädsla |
| **Swedish** | SKAM<br>BLYGSEL<br>VANÄRA<br>SKAMSENHET | SKULD<br>BROTTSLIGHET<br>SKULDMEDVETENHET | SMÄRTA<br>ONT<br>VÄRK<br>PINA<br>KVAL<br>PLÅGA<br>SORG<br>ÄNGSLAN<br>STRAFF | FÖRLÄGENHET<br>PENNINGKNIPA<br>FÖRVIRRING<br>BRYDERI<br>BESVÄR<br>HINDER<br>BETRYCK | RÄDSLA<br>FRUKTAN<br>ÄNGSLAN<br>FARHÅGA<br>RISK |
| **Tamil** | அவமானம்<br>வெட்கம் | குற்றம்<br>பிழை | வலி | மன உளைவு | భయము |
| **Telugu** | పాపం<br>అయ్యయ్యో | తప్పుచేయుట<br>మారువేషము<br>నేరము | నొప్పి<br>బాధ | ఇబ్బంది<br>చిరాకు<br>తడమటం | భయము<br>దిగులు<br>అందోళన |
| **Thai** | ความอัปยศ<br>ความอับอาย<br>หิริ<br>ความละอายใจ<br>ความอดสู<br>ความขายหน้า | ความผิด<br>ตราบาป<br>มลทิน<br>โทษโพย<br>ความรู้สึกผิด | ความเจ็บปวด<br>เจ็บ<br>ความปวด<br>ตบะ<br>ความเจ็บแสบ<br>สิ่งระคายเคือง | ความลำบากใจ<br>อุทธัจ<br>ความอึดอัดใจ<br>ความกระดาก<br>อุธัจ<br>ความตะขิดตะขวงใจ<br>ความยุ่งใจ | ความกลัว<br>ความหวาดกลัว<br>ความวิตกกังวล<br>วิตกจริต |
| **Turkish** | utanç<br>ayıp<br>yazık<br>utanma<br>yüz karası<br>leke<br>utanılacak şey | suçluluk<br>suç<br>kabahat<br>günahkârlık | ağrı<br>acı<br>sızı<br>sancı<br>ızdırap<br>eziyet<br>azap<br>dert<br>elem<br>ceza<br>emek<br>zahmet | utanma<br>sıkıntı<br>rahatsızlık<br>şaşkınlık<br>parasızlık | korku<br>korkma<br>endişe<br>kaygı<br>dehşet<br>çekinme<br>risk<br>sıkıntı<br>dert |
| **Ukranian** | сором<br>ганьба<br>ганебність<br>згуда<br>неслава<br>нага<br>нечесть | вина<br>провина<br>гріх<br>провинність<br>завина<br>караність | біль<br>страждання<br>болісність | збентеження<br>утруднення<br>вагання<br>заплутаність<br>нерішучість | страх<br>побоювання<br>зляк<br>трепет<br>обава |
| **Urdu** | شرم کی بات ہے | جرم | درد | شرمندگی | خوف |
| **Vietnamese** | điều hổ thẹn<br>điều ô nhục<br>sự xấu hổ | sự phạm tội<br>có tội | đau<br>cơn đau<br>hình phạt<br>sự đau đớn<br>nổi khổ đau<br>sự đau về xác thịt | không tự nhiên<br>sự khó chịu<br>sự ngượng ngùng<br>sự lúng túng | sợ<br>lo ngại<br>sợ hải<br>sự tôn trọng pháp luật |
| **Welsh** | cywilydd | euogrwydd | poen | embaras | ofn |
| **Yanomami** | kirii<br>kirihiwe<br>kirimai<br>kirihiprai<br>kirihou<br>no kiri thai<br>no kiri thamou<br>puhi no preami | mayo<br>no kiriai<br>no kirihiai | nini | puhipuhipe<br>sheta<br>puhi no preami<br>kiri thai<br>kiri ai | Kirii<br>kirihiwe<br>kiritimi<br>kiriri<br>kirimai<br>kirihiprai |
| **Yiddish** | שאַנד<br>ביזויען<br>בושע<br>כאַרפּע | שולד | ווייטיק | פאַרלעגנהייַט | מוירע<br>פּאַכעד<br>שרעק |

**Figure 2**: Relation between the number of words for “shame” and for “guilt” as translated by Google Translate for each of 59 languages**.** The line shows the linear regression with its 95% confidence interval.

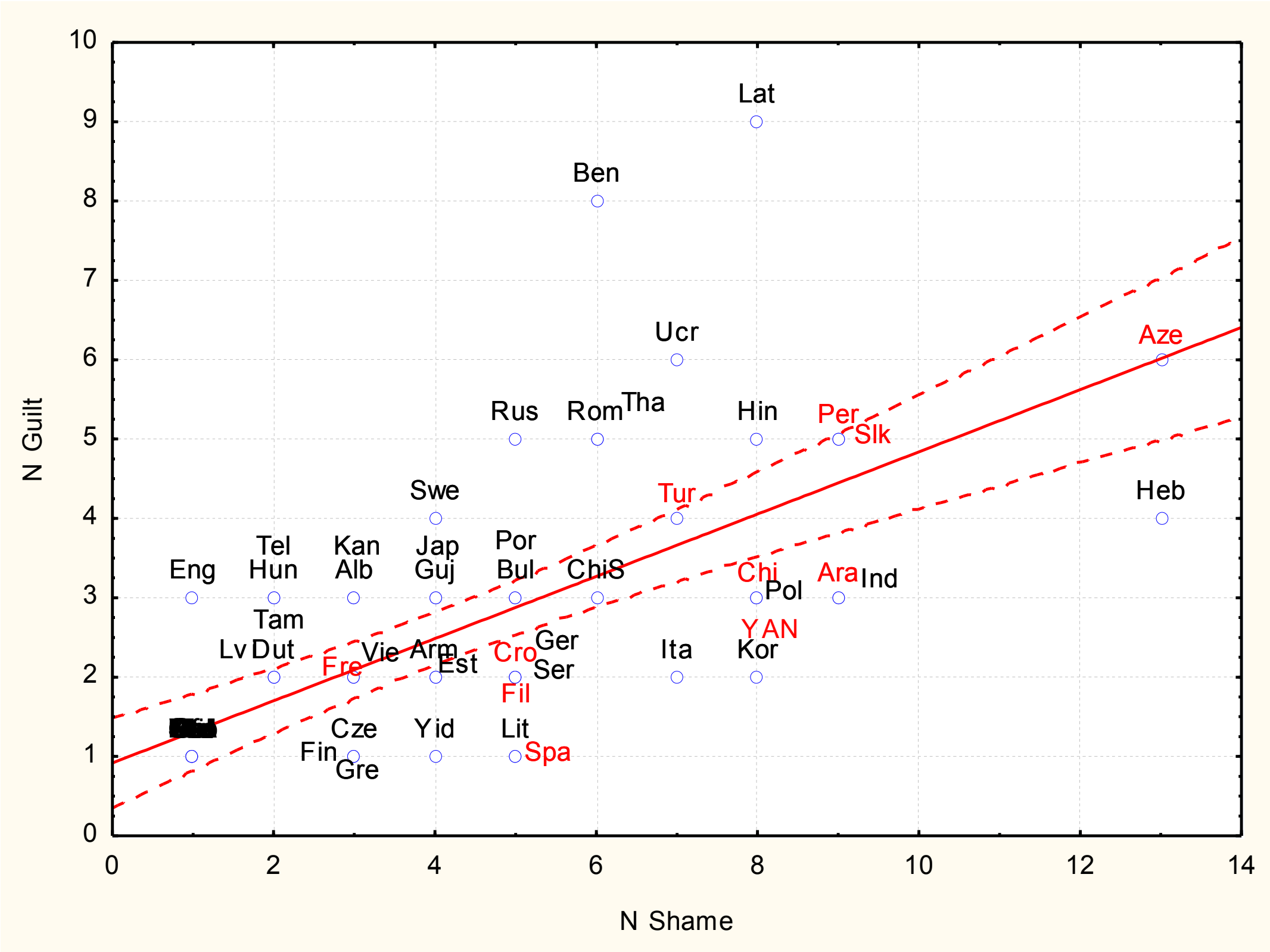

| | | | | | | | | | |
|---|---|---|---|---|---|---|---|---|---|
| Afr | **Afrikaans** | Cze | **Czech** | Hai | **Haitian** | Lit | **Lithuanian** | Spa | **Spanish** |
| Alb | **Albanian** | Dan | **Danish** | Heb | **Hebrew** | Mac | **Macedonian** | Swa | **Swahili** |
| Ara | **Arabic** | Dut | **Dutch** | Hin | **Hindi** | Mal | **Malay** | Swe | **Swedish** |
| Arm | **Armenian** | Eng | **English** | Hun | **Hungarian** | Mte | **Maltese** | Tam | **Tamil** |
| Aze | **Azerbaijani** | Est | **Estonian** | Ice | **Icelandic** | Nor | **Norwegian** | Tel | **Telugu** |
| Bas | **Basque** | Fil | **Filipino** | Ind | **Indonesian** | Per | **Persian** | Tha | **Thai** |
| Bel | **Belarusian** | Fin | **Finnish** | Iri | **Irish** | Pol | **Polish** | Tur | **Turkish** |
| Ben | **Bengali** | Fre | **French** | Ita | **Italian** | Por | **Portuguese** | Ukr | **Ukrainian** |
| Bul | **Bulgarian** | Gal | **Galician** | Jap | **Japanese** | Rom | **Romanian** | Urd | **Urdu** |
| Cat | **Catalan** | Geo | **Georgian** | Kan | **Kannada** | Rus | **Russian** | Vie | **Vietnamese** |
| ChiS | **Chinese Simple** | Ger | **German** | Kor | **Korean** | Ser | **Serbian** | Wel | **Welsh** |
| Chi | **Chinese Traditional** | Gre | **Greek** | Lat | **Latin** | Slk | **Slovak** | Yid | **Yiddish** |
| Cro | **Croatian** | Guj | **Gujarati** | Lvi | **Latvian** | Slv | **Slovenian** | Yan | **Yanomami** |

**Figure 3**: Principal component analysis built by comparing the number of words for "shame" and "guilt" and their ratio, as translated by Google Translate for each of 59 languages**.** Languages indicated in colors belong to the same language family.

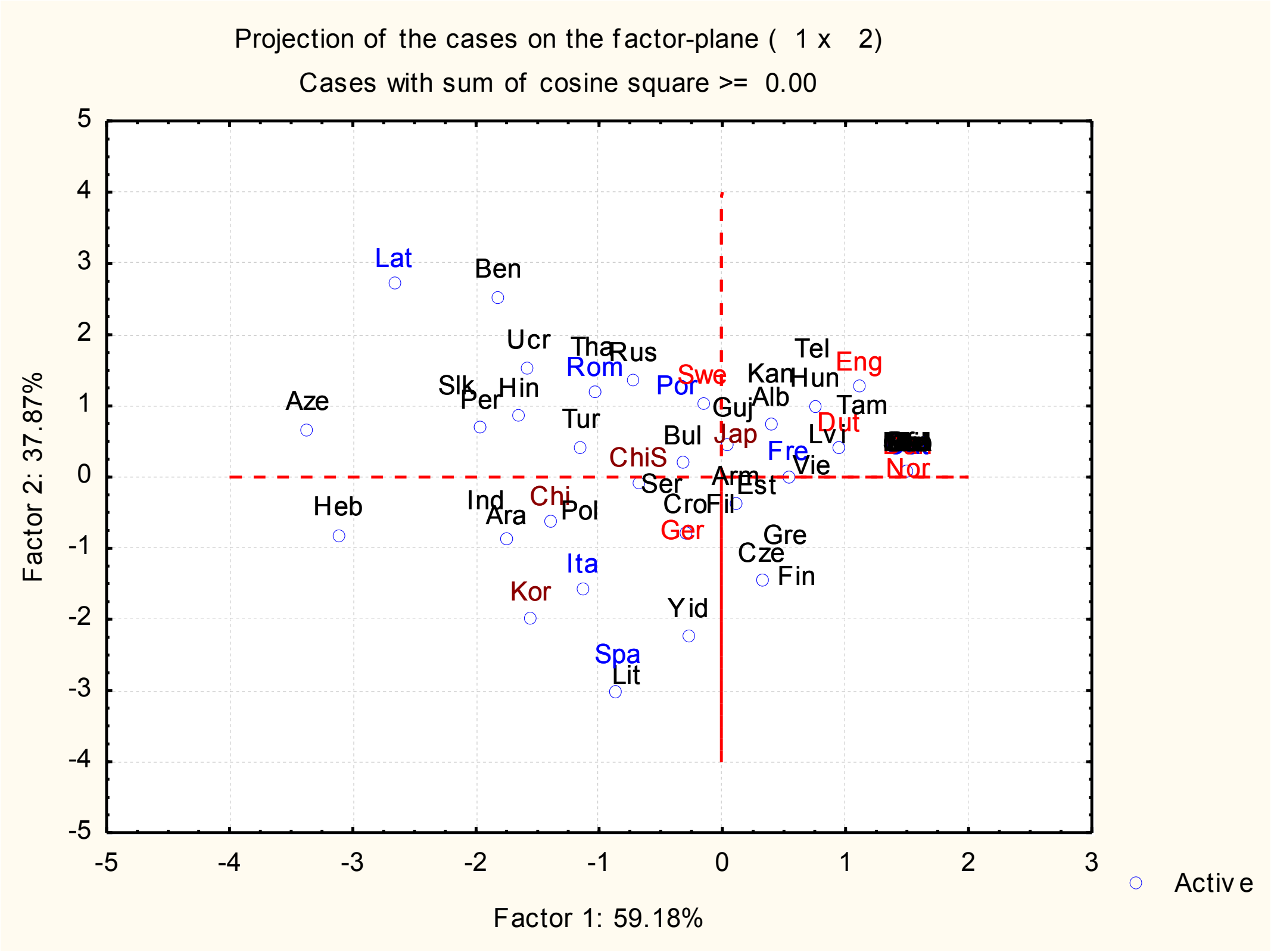

**Figure 1**: Synonyms in different languages as given by Google Translate in August 2012, except for Yanomami (31)

(ver archivo en pdf y word)